\documentclass[%
 reprint,showpacs,
 superscriptaddress,
 amsmath,amssymb,
 aps,
 prl,
]{revtex4-1}
\usepackage{longtable}
\usepackage{graphicx}
\usepackage{dcolumn}
\usepackage{bm}
\usepackage[
   bookmarks=true,
   colorlinks=true,
   hyperindex=true,
   citecolor=blue,
   linkcolor=blue,
   urlcolor=blue
]{hyperref}

\begin{document}
\title{United test of the equivalence principle at $10^{-10}$ level using mass and internal energy specified atoms}

\author{Lin Zhou}
\affiliation{State Key Laboratory of Magnetic Resonance and Atomic and Molecular Physics, Wuhan Institute of Physics and Mathematics, Chinese Academy of Sciences-Wuhan National Laboratory for Optoelectronics, Wuhan 430071, China}
\affiliation{Center for Cold Atom Physics, Chinese Academy of Sciences, Wuhan 430071, China}
\author{Chuan He}
\affiliation{State Key Laboratory of Magnetic Resonance and Atomic and Molecular Physics, Wuhan Institute of Physics and Mathematics, Chinese Academy of Sciences-Wuhan National Laboratory for Optoelectronics, Wuhan 430071, China}
\affiliation{School of Physics, University of Chinese Academy of Sciences, Beijing 100049, China}
\author{Si-Tong Yan}
\affiliation{State Key Laboratory of Magnetic Resonance and Atomic and Molecular Physics, Wuhan Institute of Physics and Mathematics, Chinese Academy of Sciences-Wuhan National Laboratory for Optoelectronics, Wuhan 430071, China}
\affiliation{School of Physics, University of Chinese Academy of Sciences, Beijing 100049, China}
\author{Xi Chen}
\affiliation{State Key Laboratory of Magnetic Resonance and Atomic and Molecular Physics, Wuhan Institute of Physics and Mathematics, Chinese Academy of Sciences-Wuhan National Laboratory for Optoelectronics, Wuhan 430071, China}
\affiliation{Center for Cold Atom Physics, Chinese Academy of Sciences, Wuhan 430071, China}
\author{Wei-Tao Duan}
\affiliation{State Key Laboratory of Magnetic Resonance and Atomic and Molecular Physics, Wuhan Institute of Physics and Mathematics, Chinese Academy of Sciences-Wuhan National Laboratory for Optoelectronics,  Wuhan 430071, China}
\affiliation{School of Physics, University of Chinese Academy of Sciences, Beijing 100049, China}
\author{Run-Dong Xu}
\affiliation{State Key Laboratory of Magnetic Resonance and Atomic and Molecular Physics, Wuhan Institute of Physics and Mathematics, Chinese Academy of Sciences-Wuhan National Laboratory for Optoelectronics, Wuhan 430071, China}
\affiliation{School of Physics, University of Chinese Academy of Sciences, Beijing 100049, China}
\author{Chao Zhou}
\affiliation{State Key Laboratory of Magnetic Resonance and Atomic and Molecular Physics, Wuhan Institute of Physics and Mathematics, Chinese Academy of Sciences-Wuhan National Laboratory for Optoelectronics, Wuhan 430071, China}
\author{Yu-Hang Ji}
\affiliation{State Key Laboratory of Magnetic Resonance and Atomic and Molecular Physics, Wuhan Institute of Physics and Mathematics, Chinese Academy of Sciences-Wuhan National Laboratory for Optoelectronics, Wuhan 430071, China}
\author{Sachin Barthwal}
\affiliation{State Key Laboratory of Magnetic Resonance and Atomic and Molecular Physics, Wuhan Institute of Physics and Mathematics, Chinese Academy of Sciences-Wuhan National Laboratory for Optoelectronics, Wuhan 430071, China}
\author{Qi Wang}
\affiliation{State Key Laboratory of Magnetic Resonance and Atomic and Molecular Physics, Wuhan Institute of Physics and Mathematics, Chinese Academy of Sciences-Wuhan National Laboratory for Optoelectronics, Wuhan 430071, China}
\affiliation{School of Physics, University of Chinese Academy of Sciences, Beijing 100049, China}
\author{Zhuo Hou}
\affiliation{State Key Laboratory of Magnetic Resonance and Atomic and Molecular Physics, Wuhan Institute of Physics and Mathematics, Chinese Academy of Sciences-Wuhan National Laboratory for Optoelectronics, Wuhan 430071, China}
\affiliation{School of Physics, University of Chinese Academy of Sciences, Beijing 100049, China}

\author{Zong-Yuan Xiong}
\affiliation{State Key Laboratory of Magnetic Resonance and Atomic and Molecular Physics, Wuhan Institute of Physics and Mathematics, Chinese Academy of Sciences-Wuhan National Laboratory for Optoelectronics, Wuhan 430071, China}
\affiliation{Center for Cold Atom Physics, Chinese Academy of Sciences, Wuhan 430071, China}
\author{Dong-Feng Gao}
\affiliation{State Key Laboratory of Magnetic Resonance and Atomic and Molecular Physics, Wuhan Institute of Physics and Mathematics, Chinese Academy of Sciences-Wuhan National Laboratory for Optoelectronics, Wuhan 430071, China}
\affiliation{Center for Cold Atom Physics, Chinese Academy of Sciences, Wuhan 430071, China}
\author{Yuan-Zhong Zhang}
\affiliation{Institute of Theoretical Physics, Chinese Academy of Sciences, Beijing 100190, China}
\author{Wei-Tou Ni}
\affiliation{State Key Laboratory of Magnetic Resonance and Atomic and Molecular Physics, Wuhan Institute of Physics and Mathematics, Chinese Academy of Sciences-Wuhan National Laboratory for Optoelectronics, Wuhan 430071, China}
\author{Jin Wang}
\email{wangjin@wipm.ac.cn}
\affiliation{State Key Laboratory of Magnetic Resonance and Atomic and Molecular Physics, Wuhan Institute of Physics and Mathematics, Chinese Academy of Sciences-Wuhan National Laboratory for Optoelectronics, Wuhan 430071, China}
\affiliation{Center for Cold Atom Physics, Chinese Academy of Sciences, Wuhan 430071, China}
\author{Ming-Sheng Zhan}
\email{mszhan@wipm.ac.cn}
\affiliation{State Key Laboratory of Magnetic Resonance and Atomic and Molecular Physics, Wuhan Institute of Physics and Mathematics, Chinese Academy of Sciences-Wuhan National Laboratory for Optoelectronics, Wuhan 430071, China}
\affiliation{Center for Cold Atom Physics, Chinese Academy of Sciences, Wuhan 430071, China}

\date{\today}

\begin{abstract}
We use both mass and internal energy specified rubidium atoms to jointly test the weak equivalence principle (WEP). We improve the four-wave double-diffraction Raman transition method (FWDR) we proposed before to select atoms with certain mass and angular momentum state, and perform dual-species atom interferometer. By combining $^{87}$Rb and $^{85}$Rb atoms with different angular momenta, we compare the differential gravitational acceleration of them, and determine the value of E\"{o}tv\"{o}s parameter, $\eta$, which measures the strength of the violation of WEP. For one case ($^{87}$Rb$|\emph{F}=1\rangle$ - $^{85}$Rb$|\emph{F}=2\rangle$),the statistical uncertainty of $\eta$ is $1.8 \times 10^{-10}$ at integration time of 8960 s. With various systematic errors correction, the final value is $\eta=(-4.4 \pm 6.7) \times 10^{-10}$. Comparing with the previous WEP test experiments using atoms, this work gives a new upper limit of WEP violation for $^{87}$Rb and $^{85}$Rb atom pairs.

\end{abstract}



\maketitle

The Einstein equivalence principle (EEP) is the heart of the general relativity (GR), it contains the weak equivalence principle (WEP). However, some new theories\cite{1} and all attempts to unify gravity with other three fundamental interactions described by the standard model\cite{2} lead to violation of the EEP. It is important to test the various aspects of EEP extensively. Atom interferometers (AIs) are basically quantum systems, which directly link quantum mechanics with GR. They meanwhile provide tools for quantum test of the WEP. Some reported experiments of WEP test using microscopic particles focus on different masses\cite{3,4,5,6,7,8}, while others only on different internal energies\cite{3,9,10} of the mass atoms.
With the test accuracy gets higher and higher, the quantum test of WEP with both mass and internal energy specified atoms becomes more significant.

The EEP has been tested with very high precision in the field of classical physics\cite{11,12,13}. In classical physics, the WEP violation is expressed by\cite{1}
\begin{equation}
m_{g}=m_{i}+\sum_{A}\eta^{A}\frac{E^{A}}{c^{2}},
\end{equation}
where, $\emph{m}_{g}$ is the gravitational mass of the test body, $\emph{m}_{i}$ is its inertial mass, $\emph{A}$ labels different interactions, $\emph{E}^{A}$ are their corresponding energies, \emph{c} is the speed of light, and $\eta^{A}$ are the WEP violation parameters. If WEP validates, then $\eta^{A}$ = 0. Generally, the WEP violation between two test bodies with mass of $m_{1}$, $m_{2}$ and acceleration of $a_{1}$, $a_{2}$, is described by the E\"{o}tv\"{o}s parameter ($\eta$) as
\begin{equation}
\eta \equiv 2\frac{|a_{1}-a_{2}|}{|a_{1}+a_{2}|}=\sum_{A}\eta^{A}(\frac{E_{1}^{A}}{m_{1}c^{2}}+\frac{E_{2}^{A}}{m_{2}c^{2}}),
\end{equation}
while the quantum test \cite{13a} of EEP is quite different from that in classical physics. In quantum formalism, the energy is described by a Hermitian operator\cite{14},
\begin{equation}
\begin{split}
\hat{H}= & mc^{2}+\frac{\hat{P}^{2}}{2m}+m\phi(\hat{Q})+\hat{H}_{int}-\hat{H}_{int}\frac{\hat{P}^{2}}{2m^{2}c^{2}} \\
         & +\hat{H}_{int}\frac{\phi(\hat{Q})}{c^{2}},
\end{split}
\end{equation}
where, \emph{m} is mass, $\hat{P}$ momentum, $\hat{Q}$ position, $\phi(\hat{Q})$ gravitational potential, and $\hat{H}_{int}$ the inertial energy operator. According to Eq. (3), the first, second and third term on the right side are related to mass, and are the classical mechanical properties of EEP. The fourth and fifth term are related to the internal energy, and are the quantum properties of EEP\cite{15}. The sixth term is the coupling between the internal energy and the gravity (gravitational time dilation)\cite{16,17,18}. If we consider the coupling between the mass and the gravitational potential, then Eq. (3) is modified by adding a term, $\phi(\hat{Q})\frac{\hat{P}^{2}}{2}$, which is similar as in Ref.\cite{19}
\begin{equation}
\begin{split}
\hat{H}= & mc^{2}+\frac{\hat{P}^{2}}{2m}+m\phi(\hat{Q})+\hat{H}_{int}-\hat{H}_{int}\frac{\hat{P}^{2}}{2m^{2}c^{2}} \\
         & +\hat{H}_{int}\frac{\phi(\hat{Q})}{c^{2}}+\phi(\hat{Q})\frac{\hat{P}^{2}}{2}.
\end{split}
\end{equation}

Obviously different test pair combination of different mass, momentum, internal energy, and so on would reveal contribution of different coupling and violation mechanics. Thus joint and comparative experiments with atom pairs of specified mass and internal energy levels are highly demanded.

A microscopic particle interferometer is naturally a quantum system. Early works of WEP test were done by neutron interferometers\cite{20,21} with an accuracy of only 10$^{-4}$. Thanks to the invention of AIs. The AI-based WEP tests exhibit much higher precision than the neutron interferometers. Fray \emph{et al}.\cite{3} and Bonnin \emph{et al}.\cite{4} used mass-different $^{87}$Rb-$^{85}$Rb atoms to test WEP with an accuracy of 10$^{-7}$ level. Schlippert \emph{et al}.\cite{5} chose $^{87}$Rb-$^{39}$K pairs to test WEP also with an accuracy of 10$^{-7}$ level. We proposed and demonstrated the four-wave double-diffraction Raman transition method (FWDR)\cite{7} for  dual-species atom interferometry, the accuracy of $^{87}$Rb-$^{85}$Rb based WEP test was improved to 10$^{-8}$ level. Barrett \emph{et al}.\cite{8} did a WEP test using $^{87}$Rb-$^{39}$K atoms in microgravity. All these efforts\cite{3,4,5,7,8} stay on the mass attribute.

On the other hand, tests with quantum attributes, like quantum statistics\cite{6}, spin\cite{9}, quantum superposition\cite{10}, and even quantum entanglement \cite{21a}, are proposed or tried experimentally. Fray \emph{et al}.\cite{3} experimented with two different \emph{F} states of $^{85}$Rb atoms. Tarallo \emph{et al}.\cite{6} used 0-spin $^{88}$Sr and half-integer-spin $^{87}$Sr atoms based on spin-gravity coupling. Duan \emph{et al}.\cite{9} used different spin orientations $(m_{F}=\pm1)$ of $^{87}$Rb atoms. Rosi \emph{et al}.\cite{10} adopted $\emph{F}=1$ and $\emph{F}=2$ of $^{87}$Rb atoms and their superposition. These works focus on the quantum internal energy properties of atoms.

\begin{figure}[htbp]
\centering
\includegraphics[width=8.0cm]{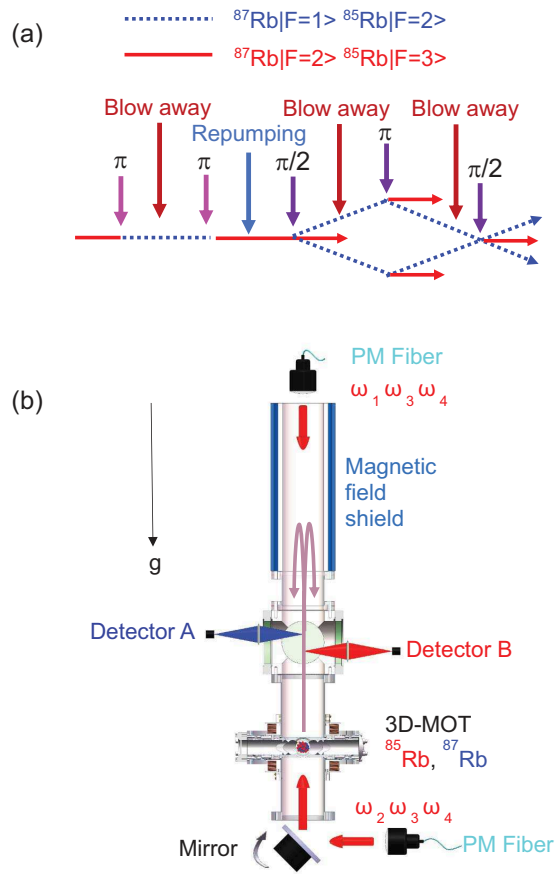}
\caption{Schematic diagram of dual-species AI and experimental setup. (a) for $^{87}$Rb$|\emph{F}=1\rangle$ and $^{85}$Rb$|\emph{F}=2\rangle$ dual-species AI, the four pulse combination ($\pi$-blow away-$\pi$-repumping) ensures that the velocity distributions of two species atoms are consistent during interference process. (b) Experimental setup.}
\label{fig:1}
\end{figure}

Now, by using the double-diffraction\cite{22,23}, dual-species, and common mode noise rejection advantages of the FWDR, we design a new scheme (Fig.\ref{fig:1}) for both mass and \emph{F} specified atom interferometer.
Taking $^{87}$Rb$|\emph{F}=1\rangle$-$^{85}$Rb$|\emph{F}=2\rangle$ dual-species AI as an example, atoms in upper states ($^{87}$Rb$|\emph{F}=2\rangle$ and $^{85}$Rb$|\emph{F}=3\rangle$) are on the non-interference path. It is necessary to remove the atoms on non-interference path and deterministically select single internal state of the atoms for the interference loop. For $^{87}$Rb$|F=1\rangle$ and $^{85}$Rb$|F=2\rangle$ atoms as shown in Fig. \ref{fig:1}(a), the $\pi$-blow away-$\pi$-repumping pulse sequence is used to perform velocity-selection and state-preparation, two blow-away pulses among $\pi/2-\pi-\pi/2$ sequence are used to clear $^{87}$Rb$|F=2\rangle$ and $^{85}$Rb$|F=3\rangle$ atoms in the interference path.

The experimental setup is shown in Fig. \ref{fig:1}(b). The three-dimensional(3D) magneto-optical trap (MOT) is in (0, 0, 1) configuration, and the four Raman lasers with frequencies of $\omega_{1}$, $\omega_{2}$, $\omega_{3}$, and $\omega_{4}$ for $^{87}$Rb and $^{85}$Rb atoms, are divided into two groups. The first group laser beams ($\omega_{1}$, $\omega_{3}$, $\omega_{4}$) pass through an 18-meter-long single-mode polarization-maintaining (PM) fiber and a beam expander, then propagate from the top window of vacuum chamber to the bottom. The second group laser beams ($\omega_{2}$, $\omega_{3}$, $\omega_{4}$) pass through a fiber, a beam expander, and a PZT-mounted mirror, then propagate from the bottom window of vacuum chamber to the top. A time division coupling method\cite{24} is used to couple multiple laser beams in one fiber.

The experimental process is briefly described as below. First, $^{87}$Rb and $^{85}$Rb atoms are pre-cooled by the two-stage two-dimensional(2D) MOT, loaded into the 3D-MOT and launched by moving molasses to form atom fountain. When the cold atom clouds go up, they are subjected velocity-selection and state-preparation process. During the subsequent rise and fall process of atoms, the FWDR operation is used to implement the dual-species AI. The free evolution time among $\pi/2-\pi-\pi/2$ Raman pulses, \emph{T}, is carefully adjusted to minimize the ellipse fitting error of data\cite{25,26}. After the interference, the atom clouds continue to fall through the detection zone, and the fluorescence of $^{87}$Rb and $^{85}$Rb atoms excited by the resonant probe beams are collected by detectors A and B, respectively. To minimize the error due to probe position and detectors, we repeat each measurement by changing $^{87}$Rb and $^{85}$Rb atoms to detectors A and B. The atom interference fringes are obtained by recording fluorescence signal verse the phase of the second $\pi/2$ Raman pulses. When the detected population data of $^{87}$Rb and $^{85}$Rb atoms are taken as the \emph{x}-axis and \emph{y}-axis, respectively, we obtain an ellipse composed of scattered data points. The fitting phase of the ellipse corresponds to the absolute difference of gravity measured by the dual-species AI. The statistical uncertainty of the WEP violation coefficient, $\eta$, can be evaluated by performing Allan deviation on a large number of differential-gravity data.

To be one example, the joint WEP test is performed firstly for atom pair of $^{87}$Rb$|\emph{F}=1, m_{F}=0\rangle$-$^{85}$Rb$|\emph{F}=2, m_{F}=0\rangle$. We use the interference scheme shown in Fig. \ref{fig:1}(a) to achieve gravity differential-measurement.

\begin{figure}[htbp]
\centering
\includegraphics[width=8.4cm]{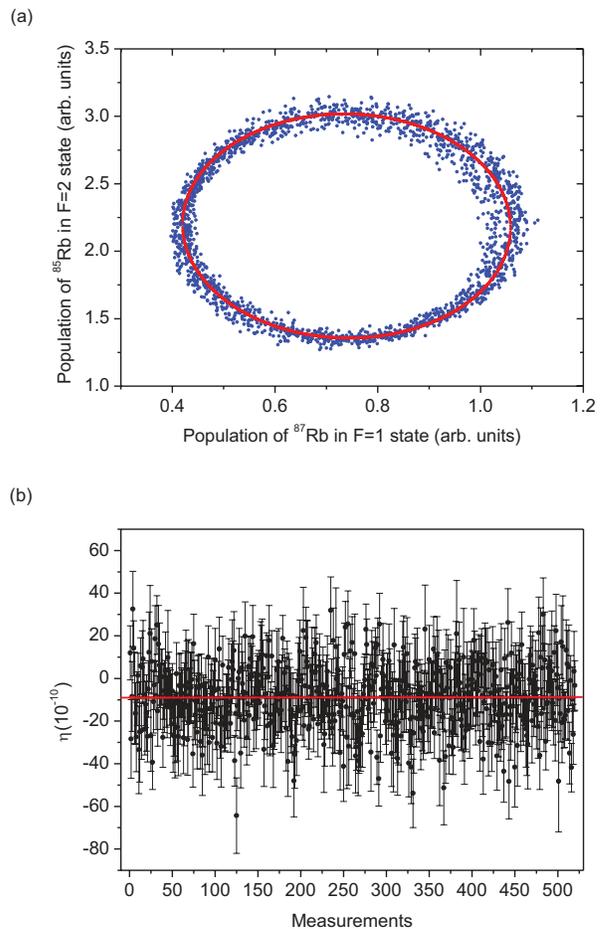}
\caption{Experimental data. (a) Elliptic signals based on the population of $^{87}$Rb$|\emph{F}=1, m_{F}=0\rangle$ and $^{85}$Rb$|\emph{F}=2, m_{F}=0\rangle$ atoms, the blue dots are 2000 experimental data, and solid red line is the fitted ellipse. (b) Experimentally measured $\eta$ values, where the error corresponding to effective wave vector is corrected, and the average value of 520 measurements is $\eta=-8.9\times10^{-10}$ as showing in red line.}
\label{fig:2}
\end{figure}

\begin{figure}[htbp]
\centering
\includegraphics[width=8.0cm]{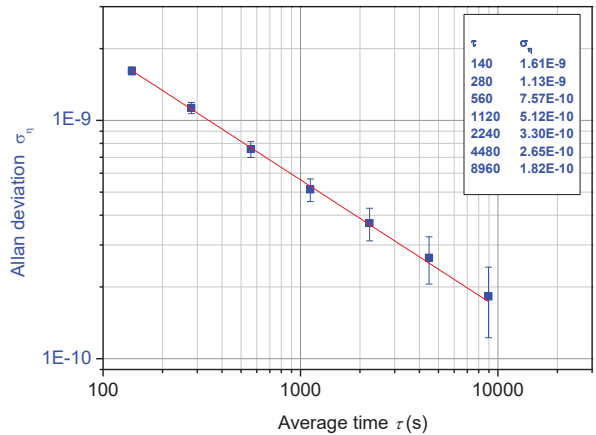}
\caption{Allan deviation of the uncertainty of $\eta$ measurements. The blue squares are measurements based on $^{87}$Rb$|\emph{F}=1\rangle$-$^{85}$Rb$|\emph{F}=2\rangle$ atoms, and the uncertainty is $1.8\times10^{-10}$ at integration time of 8960 s. The red line is fitted curve.}
\label{fig:3}
\end{figure}

The experimental data of $^{87}$Rb$|\emph{F}=1, m_{F}=0\rangle$ and $^{85}$Rb$|\emph{F}=2, m_{F}=0\rangle$ atoms is shown in Fig. \ref{fig:2}. The elliptic signal in Fig. \ref{fig:2}(a) is obtained by the population of $^{87}$Rb$|\emph{F}=1, m_{F}=0\rangle$ and $^{85}$Rb$|\emph{F}=2, m_{F}=0\rangle$ atoms, each ellipse consists of 40 measurement data points (with measurement time of 140 s, and the free evolution time \emph{T}=203 ms). Fig. \ref{fig:2}(b) shows 520 measurements of differential-data between the gravitational acceleration of $^{87}$Rb$|\emph{F}=1, m_{F}=0\rangle$ and $^{85}$Rb$|\emph{F}=2, m_{F}=0\rangle$ atoms, where each data point is given by an ellipse fitting. The average value of the 520 measurements is  $49426.6\times10^{-10}$,
and the error corresponding to effective wave vector is $49435.5\times10^{-10}$, thus the corrected value of $\eta$ is $-8.9\times10^{-10}$. The Allan deviation given by the data in Fig. \ref{fig:2}(b) is shown in Fig. \ref{fig:3}. The Allan deviation gives an uncertainty of $1.8\times10^{-10}$ of $\eta$ measurement based on $^{87}$Rb$|\emph{F}=1\rangle$ and $^{85}$Rb$|\emph{F}=2\rangle$ atoms at integration time of 8960 s.

\begin{table}[htbp]
\caption{\textbf{The error budget}}
{\begin{tabular}{@{}lcc@{}}
\toprule
Parameters & $\eta$ ($\times10^{-10}$)& Uncertainty($\times10^{-10}$)  \\
\colrule
Experimental data &  49426.6 &  1.8  \\
\colrule
Effective wave vector & 49435.5 & 0.5  \\
Detector difference & 1.0 & 2.5  \\
Coriolis effect &  0 &  2.0  \\
Wave-front aberration &  0 &  5.0 \\
ac Stark shift &  0 &  0.6  \\
Quadratic Zeeman shift &  0 &  2.0  \\
Gravity gradient &  -5.5 &  1.2  \\
Others &  0 &  1.0  \\
\colrule
Total & -4.4 & 6.7 \\
\botrule
\end{tabular} \label{ta1}}
\end{table}

The uncertainty of the wave vector is $0.5\times10^{-10}$, which is caused by the uncertainty of the local gravity.
The error due to the difference of probe position and detectors is evaluated as $1.0\times10^{-10}$ with an uncertainty of $2.5\times10^{-10}$.
We carry out rotation compensation on the mirror of Raman beams, and decrease the measurement uncertainty caused by the Coriolis effect to $2.0\times10^{-10}$. The upper limit of the wave-front aberration error analyzed from the actual experimental parameter is $5.0\times10^{-10}$. The ac Stark shift uncertainty evaluated by modulation experiments is $0.6\times10^{-10}$. The error corresponding to the quadratic Zeeman shift is $2.0\times10^{-10}$. The error corresponding to gravity gradient is $(-5.5\pm1.2)\times10^{-10}$.

There are other systematic error terms, the time-dependent fluctuation of absolute gravity (solid tide) is less than $1.2\times10^{-12}$, laser frequency stability is better than 1 MHz, the chirp rate of Raman laser frequency is better than 1 Hz/s, the laser beam pointing accuracy is controlled to be better than 0.1 mrad, and the time control accuracy is better than 1 ns. Thanks to the common-mode noise-suppression of the FWDR method, the influence of these factors on the measurement uncertainty is less than $1.0\times10^{-13}$. The higher-order systematic errors\cite{27}, such as collision shift and wavelength fluctuations, are suppressed to less than $1.0\times10^{-11}$. The contribution of all these systematic error terms is no more than $1.0\times10^{-10}$. The total systematic error of gravity differential measurement using $^{87}$Rb$|\emph{F}=1\rangle$-$^{85}$Rb$|\emph{F}=2\rangle$ dual-species AI is $(-4.4\pm6.7)\times10^{-10}$. The main systematic error budgets are shown in Table \ref{ta1}.

The reported classical tests of EP usually use large mass macroscopic objects\cite{11,12,13,28}, the accuracy of these tests is at $10^{-13} \sim 10^{-15}$ level. Atomic clocks\cite{29} are also candidates for EP test. The test for a new kind of EP was demonstrated with the extended rotating bodies at $10^{-7}$ level\cite{30}. The best accuracy of the mass test of WEP using atoms is at 10$^{-8}$ level\cite{7}, and the current accuracy of quantum test of WEP using single species atoms is at 10$^{-9}$ level\cite{10}. Compared with all the previous experiments, this work jointly tests the WEP using both mass and $\emph{F}$ specified dual-species AI. The measurement results show that the WEP of the microscopic particles is still validated at the accuracy of $6.7\times10^{-10}$.

In summary, we improved the FWDR dual-species atom interferometry and performed a quantum test of WEP in small mass microscopic particle field using mass specified and internal energy specified atoms. The experimental result gives a new upper limit for microscopic particle based WEP violation. Our experiments show that the statistical Allan deviation for other combinations of different mass and \emph{F} state rubidium atoms, $^{87}$Rb$|\emph{F}=2\rangle$-$^{85}$Rb$|\emph{F}=3\rangle$, $^{87}$Rb$|\emph{F}=1\rangle$-$^{85}$Rb$|\emph{F}=3\rangle$, and $^{87}$Rb$|\emph{F}=2\rangle$-$^{85}$Rb$|\emph{F}=2\rangle$, are all at 10$^{-10}$ level. The detailed systematic error budget for these combinations is under way.

This work was supported by National Key Research and Development Program of China under Grant No. 2016YFA0302002, National Natural Science Foundation of China under Grant Nos. 91536221, 91736311, 11574354 and 11204354, Strategic Priority Research Program of the Chinese Academy of Sciences under grant Nos. XDB21010100, XDB23030100, and Youth Innovation Promotion Association of the Chinese Academy of Sciences.

\end{document}